# Semi-Empirical Neutrino and Quark Mixing Angle Broken 4-Color Geometric Symmetry

E. M. Lipmanov
40 Wallingford Road # 272, Brighton MA 02135, USA

## Abstract

Two semi-empirical ideas are guiding the present research: 1) one neutrino and N color quark triple mixing angles have second physical meaning of vector direction angles in euclidean 3-space, and 2) the difference between neutrino and quark mixing in the SM weak interactions is mostly conditioned by known fundamental difference of these particles by numbers of color degree of freedom. As main physical inference, the elementary particle mixing angles obey a novel phenomenological quark-neutrino (N+1)-color and 3 mixing angle symmetric positive definite quadratic Pythagorean equation in terms of double mixing angles. This equation predicts immediately that two large solar and atmospheric angles must be accompanied by one small not zero theta-13 neutrino angle in convincing agreement with recent new theta-13 experimental data. Benchmark flavor mixing pattern – united bimaximal neutrino and zero quark ones – follows from solutions of the primary equation without free parameters. Agreement with known experimental data follows only when the number of quark colors is equal to the number of particle flavors, N = 3. Together with closely related to the fine structure constant one empirical small parameter, that benchmark pattern determines realistic quark and neutrino mixing angles in good agreement with data and especially accurate for quark Cabibbo and neutrino solar angles. Coming accurate neutrino mixing angle data are important for testing the Pythagorean equation.

## 1. Introduction.

Without well established flavor theory[1], new suggestive empirical quark and lepton flavor regularities cannot be excluded and can be discovered (sometimes by serendipity, as known from physics history). By definition, semi-empirical flavor mixing phenomenology is a *system* of regularities and appropriate generalizations that have predictive power and may be justified as possible empirical basis for new flavor theory. Though different from the ongoing major symmetry-quest, the search in this paper for a consistent semi-empirical phenomenology of united neutrino and quark mixing patterns modestly follows the long time traditions of problem solving in frontier physics as mainly experimental science.

Today flavor physics seems fundamental, but not complete. It reminds the early stage of quantum mechanics[2].

## 2. Basic Pythagorean equation for particle mixing angles

As original hypothesis, 1-color neutrino and N-color quark mixing angle triplicates are represented by 3-vectors in euclidean 3-dimensional mixing-angle-space with main physical inference of a new phenomenological quark-neutrino (N+1)-color and 3 mixing angle symmetric positive definite quadratic Pythagorean equation,

$$\cos^2 2\theta^k_{12} + \cos^2 2\theta^k_{23} + \cos^2 2\theta^k_{13} = 1. \qquad (1)$$

Superscript index 'k' in (1) denotes (N+1) particle colors: '0' is for neutrinos and '1−N' are for N quark colors. As an important feature, Eq. (1) is expressed in terms of double[3] ($2\theta^k_{ij}$) particle mixing angles.

Eq. (1) is characterized by two different symmetry types.

1) Double mixing angle symmetry under permutation of the three lower indexes 12, 23 and 13. The reason for that symmetry with three mixing angles is that the neutrino and color-quarks mixing angles have meaning of direction angles of vectors (their lengths do not have physical meaning) in a 3-dimensional euclidean space formally identical to the regular macroscopic 3-space. Considering an orthogonal coordinate system (**X, Y, Z**) in this space, the direction angles of vectors $\mathbf{N}^k$ are expressed through neutrino double mixing angles

$$\mathbf{N}^k = (\cos 2\theta^k_{12}, \cos 2\theta^k_{23}, \cos 2\theta^k_{13}). \qquad (2)$$

---

[1] To date there is no such theory, though a grate many of flavor symmetry models are published in the literature for decades. Maybe some of them will pave the way for true theory, but of cause there is no guarantee of that. It seems that semi-empirical flavor phenomenology based on empirical regularities may be interesting and helpful.

[2] As known, based on Plank constant semi-empirical quantum phenomenology preceded the new paradigm of quantum theory discovered by Heisenberg, Schrödinger, Born and Dirac.

[3] Other choices in '$n\theta^k_{ij}$' with $n \neq 2$, including the trivial coefficient $n = 1$, are excluded by experimental data.

For visualization, consider a geometric cube-model with three normalized solutions of Eq.(1) for vector $\mathbf{N}^k$:

$$(1\ 0\ 0) = \mathbf{a},\ (0\ 1\ 0) = \mathbf{b},\ (0\ 0\ 1) = \mathbf{c} \qquad (3)$$

are the three cube edges, and the solution

$$\mathbf{d} = (1\ 1\ 1)\ /\sqrt{3} \qquad (4)$$

is its diagonal (arbitrary units of length). The three cube edges **a**, **b**, **c** and coordinate axes **X**, **Y, Z** are respectively chosen parallel.

At three mixing angles (and only at three) the numbers of particle flavors and mixing angles are equal. So, the geometric space-like symmetry of Eq. (1) means equal numbers of particle flavors, mixing angles and space dimensions (by definition).

2) Quark-neutrino symmetry under permutation of the (N+1) upper indexes between neutrino and N quark colors. This symmetry requires equal numbers of quark and neutrino flavors: $n^{(q)}_f = n^{(\nu)}_f$.

There is only one solution of Eq. (1) that is universal and completely symmetric under permutation of lower mixing angle indexes 'ij' and upper neutrino and quark color indexes 'k',

$$\cos^2 2\theta^k_{ij} = 1/3,\ \ \theta^k_{ij} \cong 27.4^{\circ},\ \ ij = 12, 23, 13;\ \ k = 0 \ldots N. \qquad (5)$$

Solution (5) is determined only by the number of mixing angles in (1) and is independent of the number N of quark colors. By geometric visualization, the double angels of solution (2) are equal those of a diagonal in a cube[4].

Well known experimental data on particle mixing angles [2] show that solution (5) is in sharp disagreement with both neutrino and quark mixing angles at low energies. But though (5) looks a nonphysical solution, it should be useful in the research since it unites color quark and neutrino mixing angles and displays full underlying symmetry of Eq. (1).

The neutrino equation, k = 0,

---

[4] (**ad**) = (**bd**) = (**cd**) $\cong \cos 54.7^{\circ}$.

$$\cos^2 2\theta^{\nu}_{12} + \cos^2 2\theta^{\nu}_{23} + \cos^2 2\theta^{\nu}_{13} = 1, \qquad (6)$$

was proposed in [1] for explanation of new T2K indications [3] on relatively large theta-13 angle by its space-like geometric symmetry relation to empirically known two large solar and atmospheric angles. Eq. (1) is factually a generalization of neutrino Eq. (6).

In contrast to neutrinos, equation (1) for quarks is in strong disagreement with experimental data. New low energy white-color-quark mixing angles $\theta^{q}_{ij}$ are defined in Sec. 3 by equation

$$\cos^2 2\theta^{q}_{12} + \cos^2 2\theta^{q}_{23} + \cos^2 2\theta^{q}_{13} = 3 \qquad (7)$$

for leading approximation benchmark quark mixing. Quark colors are hidden in the color-symmetric quantities $\cos^2 (2\theta^{q}_{ij})$, they are color symmetric in agreement with QCD. Eq. (7) means zero quark mixing angles in appropriate approximate agreement with CKM data [2].

Equations (6) and (7) define the concept of benchmark particle mixing pattern without free parameters that include united bimaximal neutrino (Sec.2) and zero quark (Sec.3) mixing patterns. It is that benchmark[5] mixing pattern that gives meaning to neutrino and quark realistic flavor mixing angles as deviations from the primary level by one small universal, probably dynamical, empirical parameter.

The united quark-neutrino benchmark mixing pattern is considered zero approximation of small empirical epsilon-parameter,

$$\varepsilon \cong 0.082085 \cong \exp(-2.5). \qquad (8)$$

This parameter is close (~ 4%) to the dimensionless-made elemental electric charge

$$e = \sqrt{\alpha} \cong 0.0854 \qquad (9)$$

where $\alpha$ is the fine structure constant at $q^2 = 0$.

Both parameters[6] (8) and (9) are independently used below for data fitting with close results.

In Sec. 2, neutrino mixing angles are obtained from the solutions of Eq. (1) by spontaneous geometric symmetry violation (benchmark pattern) and epsilon-parametrization (realistic pattern). In Sec. 3, Eq. (7) is derived from Eq. (1), and realistic small quark mixing angles are obtained. In Sec. 4, three complementarity relations between neutrino and quark mixing angles are obtained. In Sec. 5, relation between epsilon-parametrization and the special concept of benchmark is illuminated. Sec. 6 contains discussions of some results.

---

[5] The important concept of benchmark was virtually introduced in theoretical physics by Newton in his laws of classical mechanics: realistic particle motion is a deviation from primary benchmark inertial motion (no parameters, geometric symmetry) caused by forces (free dynamical parameters, symmetry violation).

[6] The preference is for the epsilon-parameter [6].

## 3. Neutrino part of benchmark mixing

Only two of the three neutrino mixing angles in Eq. (6) are independed. If one angle is small ($<< \pi/2$), both other two angles must be large; in reverse, if two angles are large, the third one must be small. Zero, two or three small mixing angles are forbidden by geometry. It is exactly what is observed experimentally for neutrino mixing pattern in contrast to the quark one.

A symmetric set of asymmetric mixing angle solutions of Eq. (6) without nontrivial parameters other than '0' and '1' is given by

$$(\cos^2 2\theta^{\nu}_{12}, \cos^2 2\theta^{\nu}_{23}, \cos^2 2\theta^{\nu}_{13}) = (0, 0, 1) = \mathbf{c},$$
$$(\cos^2 2\theta^{\nu}_{12}, \cos^2 2\theta^{\nu}_{23}, \cos^2 2\theta^{\nu}_{13}) = (0, 1, 0) = \mathbf{b},$$
$$(\cos^2 2\theta^{\nu}_{12}, \cos^2 2\theta^{\nu}_{23}, \cos^2 2\theta^{\nu}_{13}) = (1, 0, 0) = \mathbf{a}. \quad (10)$$

By summing up all three asymmetric solutions the universal symmetric solution (5) follows

$$(\mathbf{a} + \mathbf{b} + \mathbf{c}) = (3\cos^2 2\theta^{\nu}_{12},\ 3\cos^2 2\theta^{\nu}_{12},\ 3\cos^2 2\theta^{\nu}_{12}) = (1, 1, 1). \quad (11)$$

In terms of the geometric cube-model, every solution in (10) represents one edge of the cube **a**, **b** or **c**, and their sum in Eq. (11) is a diagonal of the cube, see footnote 4.

The set of solutions (10) is a symmetric one until the mixing angles are physically identified by experimental data. Physical identification of mixing angles by known experimental data [2] singles out the upper line solution in the set (10),

$$\cos^2(2\theta^{\nu}_{12}) = 0,\ \cos^2(2\theta^{\nu}_{23}) = 0,\ \cos^2(2\theta^{\nu}_{13}) = 1;\ \theta^{\nu}_{12} = \theta^{\nu}_{23} = \pi/4,\ \theta^{\nu}_{13} = 0. \quad (12)$$

Its physical meaning is two $\theta^{\nu}_{12}$ and $\theta^{\nu}_{23}$ maximal and one $\theta^{\nu}_{13}$ zero mixing angles of benchmark neutrino mixing pattern. This singling out of solution (12) by condition of approximate agreement with data is a 'spontaneous symmetry violation'[7] of Eq. (6), or set (10), symmetry. In geometric terms, this spontaneous symmetry violation singles out the edge **c** as neutrino mixing vector $\mathbf{N}^{\nu} = \mathbf{c} = (0, 0, 1)$ in the cube.

---

[7] It is by the original meaning of the term "spontaneous symmetry violation".

An appropriate comment is due. The used above term 'spontaneous symmetry violation' is different from the common one in the electroweak theory that is related to not zero vacuum expectation value of a scalar field. In semi-empirical phenomenology spontaneous symmetry violation should be related to its original meaning that is 'singling out one solution from a symmetric set of nonsymmetrical solutions of a symmetric equation'.

Unitary mixing matrix with angles (12) is the bimaximal neutrino matrix,

$$\begin{pmatrix} 1/\sqrt{2} & 1/\sqrt{2} & 0 \\ -1/2 & 1/2 & 1/\sqrt{2} \\ 1/2 & -1/2 & 1/\sqrt{2} \end{pmatrix}_{\nu} . \qquad (13)$$

This widely discussed in the literature neutrino mixing matrix approximately describes empirical neutrino mixing data [5].

Bimaximal mixing pattern appears an appropriate starting level (benchmark) for realistic neutrino mixing angles as small deviations from benchmark by epsilon-parameterization, see Appendix.

## 4. Quark part of benchmark mixing

Starting with Eq. (1) and summing up the N color-quark equations we get

$$\sum\nolimits^{c}_{(1\to N)} (\cos^2 2\theta^c_{12} + \cos^2 2\theta^c_{23} + \cos^2 2\theta^c_{13}) = N. \qquad (14)$$

But individual color quark mixing angles $\theta^c_{ij}$ in the weak interactions are not observable at low energies. New color-symmetric 'white' quark mixing angles $\theta^q_{ij}$ are defined by relations

$$\sum\nolimits^{c} (\cos^2 2\theta^c_{ij}) \equiv \cos^2 2\theta^q_{ij} \leq 1 \qquad (15)$$

via sum over N quark colors. Eq. (14) is a semi-empirical relation that leads from the color-quark mixing angles $\theta^c_{ij}$, that obey the primary Eq, (1), to realistic quark mixing angles $\theta^q_{ij}$ at benchmark approximation in agreement with low energy quark mixing angles in the weak interaction Lagrangian[8].

Equation (14) with definition (15) determines one white-color quark mixing angle equation

$$\cos^2 2\theta^q_{12} + \cos^2 2\theta^q_{23} + \cos^2 2\theta^q_{13} = N. \qquad (16)$$

[8] The physical meaning of the color quark mixing angles $\theta^c_{i\,j}$, as indicated in [12], may be related to some high energy scale.

Note, in an imagined world without strong color QCD interactions the quark colors would be observable and identical, $\cos^2 2\theta^q_{ij} = N\cos^2 2\theta^c_{ij}$. Then Eq.(16) for individual colors[9] would coincide with neutrino equation (3) Thus quark QCD interactions are important implied conditions in the definition (15) and Eq. (16).

A confirmation of the definition (15) consistency for quark mixing angles is that together with Eq. (16) it means supported by data condition of equal numbers of quark colors and quark flavors. Indeed, the positive-definite Eq. (16) is a restriction on the number of colors. It allows only a few choices for the number of quark colors $N \leq 3$, $N = 0, 1, 2$ or 3. From known data on CKM quark mixing angles [2], three values $N = 0, 1, 2$ are excluded and only one left $N = 3$ – three colors. With three colors Eq. (16) coincides with Eq. (7).

For completeness, consider another (the most detailed) derivation of Eq. (7) especially because it underlines the relation between quark and neutrino solutions of Eq. (1). Each of the N color quark equations in (1) evidently has a symmetric set of three asymmetric solution of the neutrino type (10):

$$(\cos^2 2\theta^c_{12}, \cos^2 2\theta^c_{23}, \cos^2 2\theta^c_{13}) = (0, 0, 1) = \mathbf{c},$$
$$(\cos^2 2\theta^c_{12}, \cos^2 2\theta^c_{23}, \cos^2 2\theta^c_{13}) = (0, 1, 0) = \mathbf{b},$$
$$(\cos^2 2\theta^c_{12}, \cos^2 2\theta^c_{23}, \cos^2 2\theta^c_{13}) = (1, 0, 0) = \mathbf{a}. \qquad (17)$$

Summing up all 9N relations[10] in Eq. (17) and using the definition (15) results in Eq.(7) (for N = 3) and empirical benchmark quark mixing with equal mixing angles $\theta^q_{ij} = 0$. It happens only in case of mentioned symmetry between all three cube edges **a, b** and **c** – not broken geometric rotational symmetry around the diagonal of the cube.

Thus there are two additional symmetries of the quark mixing pattern at benchmark point that are absent in neutrino benchmark mixing pattern – quark three-color symmetry and symmetry between three cube edges.

---

[9] It should be noticed that even at absence of QCD interactions the quark mixing would be not of the same type as neutrino one because of spontaneous violation of the cube-edge symmetry only in neutrino case, compare (21) for quarks and (8)-(9) for neutrinos.

11 Notice the physical meaning of these 9N quark relations (17): 3N equations in each row are related to 3 mixing angles (three flavors) for N colors 'c', and the 3 rows in (17) correspond to three edges (**a, b, c**) of the cube for each color in the geometric visualization model, 3x3N = 9N.

An important result of the derivations of Eq. (7): experimental data on quark mixing angles in considered phenomenology definitely require equal numbers of quark colors and quark mixing angles and so − equal numbers of quark colors and quark flavors.

The final one equation for the mixing angles $\theta^q_{ij}$ at benchmark point is a sum over all 27 terms in (17) (at N = 3) and definition (15). It coincides with Eq. (7),

$$\cos^2 2\theta^q_{12} + \cos^2 2\theta^q_{23} + \cos^2 2\theta^q_{13} = 3. \tag{7'}$$

Eq. (7) has only one solution,

$$\cos^2 2\theta^q_{12} = \cos^2 2\theta^q_{23} = \cos^2 2\,\theta^q_{13} = 1, \tag{18}$$

$$\theta^q_{12} = \theta^q_{23} = \theta^q_{13} = 0. \tag{19}$$

Unitary quark mixing matrix with angles (19) is a unit matrix. It describes benchmark quark mixing angles (zero ε-approximation) that approximately reproduce the small realistic CKM mixing angles [2].

## 5. Conclusions

Quark color involvement in flavor mixing is in the spirit of united three elementary particle forces. Guided by the semi-empirical ideas that neutrino and every one of N color-quark mixing angles are also direction angles of a vector in euclidean 3-spacet and stimulated by the experimental indications on unexpectedly large theta-13 neutrino mixing angle, one quadratic positive-definite double mixing angle quark-neutrino (N + 1)-color symmetric Pythagorean Eq. (1) is investigated.

The main new results of the present research are

**{1}** Basic concept of 'benchmark' in particle mixing phenomenology, as the level of reference without free parameters, is determined by symmetric primary Eq. (1), and its spontaneous violation in the special case of neutrinos.

A central position in the present research occupies the universal particle mixing angle solution (2) of Eq. (1) that is visualized in the geometric cube-model by a symmetric sum of three solutions, marked by three cube-edges **a, b** and **c** and is represents by the cube-diagonal**, d = (a + b + c).** The benchmark solution for the neutrino mixing angles (bimaximal mixing pattern) is obtained from this universal solution by spontaneous (**a, b, c**)-symmetry violation that singles out the cube-edge **c**. In contrast to neutrinos, the individual color-quark mixing would be exactly the cube-edge symmetric solution (2), but is not observable at low energies because of color confinement by strong QCD interactions. A successful relation between the individual color-quark mixing angles and the observable at low energies color-white angles is introduced, it determines zero mixing angle quark solution that is a leading approximation to the CKM data.

**{2}** Symmetry of Pythagorean neutrino Eq. (6) near-perfectly describes the pattern of empirical neutrino mixing angles. It allows two large and one small angle and forbids zero, two or three small angles in contrast to three small quark mixing angles.

**{3}** Known large solar and atmospheric neutrino mixing angles predict not very small reactor theta-13 angle in conformity with new experimental indications. The reasons of why the neutrino mixing angle theta-13 is predicted as relatively large are the not very small deviations from maximal values of the two large neutrino mixing angles, most importantly the solar one.

**{4}** Analogy between quark and neutrino mixing angles do exist, but only at the initial level of individual color quark mixing angles and not-violated geometric cube-edge symmetry.

**{5}** Essential differences between quark and neutrino mixing at benchmark are two additional symmetries in the quark pattern that are missing in the neutrino one – 3-color symmetry and symmetry between the three cube-edges (**a, b, c**) in quark mixing pattern.

**{6}** Widely discussed in the literature bimaximal neutrino and zero quark mixing patterns are predicted by the symmetric Eq. (1) as substantially connected by origin benchmark mixing patterns without free parameters.

**{7}** Data supported quark benchmark mixing pattern follows from Eq. (1) only at three colors $N = 3$ by new definition of the three low energy white mixing angles.

**{8}** Realistic neutrino and quark mixing angles in good agreement with data are obtained as deviations from primary benchmark values by one small related to the fine structure constant parameter $\varepsilon$ or $\sqrt{\alpha}$, see Appendix.

**{9}** A new universal small empirical parameters (8) is well motivated in considered particle mixing semi-empirical phenomenology.

Benchmark neutrino and quark mixing angles are determined by *symmetry* of Eq. (1) and its spontaneous violation (only in neutrino case), while the realistic low energy deviations of the mixing angles from benchmark values are additionally determined by new *dynamics* represented in the considered semi-empirical phenomenology by one small ε-parameter.

The main justification of the small parameter is that it works well in the considered phenomenology. Conformable and simple ε-parametrizations for all six quark and neutrino low energy mixing angles are factually in fair agreement with experimental data values. An interesting feature of the ε-parameter is that it may be related to all three basic SM interactions by simple connection to GUT-constant, $\alpha_{GUT} \cong \varepsilon / 2 \cong 1/25$ [15].

# Appendix

## 1. Theta-13 from Pythagorean equation and solar and atmospheric data

Neutrino Pythagorean Eq. (6) defines theta-13 through the experimental solar and atmospheric angles

$$\sin^2 2\theta_{13} = \cos^2 2\theta_{12} + \cos^2 2\theta_{23}. \quad (A1)$$

Using recent results of the global 3ν neutrino oscillation analysis [**9**] for solar angle $\theta_{12}$ at 2**σ**,

$$\sin^2 \theta_{12} = 0.275 - 0.342 \quad (A2)$$

and maximum value for the atmospheric angle $\theta_{23} = 45\,^{o}$, which is indicated by T2K data [10] and agrees with the global analysis [9] at 1**σ**, one gets

$$\theta_{13} = 9.2\,^{o} - 13.3\,^{o}. \quad (A3)$$

As important results, the ranges (9) of reactor angle from Eq. (2) are in very good agreement with latest T2K $\nu_\mu \to \nu_e$ appearance data [1] and compatible with independent global analysis of reactor angle $\theta_{13}$ ranges [**9**] at 1**σ** (normal or inverted hierarchy)

$$\sin^2 \theta_{13} = 0.0216 - 0.0260, \quad \theta_{13} = 8.6\,^{o} - 9.3\,^{o}. \quad (A4)$$

Further verification of this compatibility by coming accurate neutrino mixing angle data will be an important test of the neutrino Pythagorean equation.

## 2. Parametrization of realistic neutrino mixing angles

Bimaximal neutrino mixing is considered benchmark for realistic neutrino mixing angles. The latter are shifted from benchmark values by small empirical ε-parameter and introduced earlier [6] universal exponential positive function f(y),

$$f(y) = |\, y \exp(y) \,|, \quad (A5)$$

with 'y' a power function of the parameter. This combination of two conditions (8), or (9), and (A5) is the definition of the term 'epsilon-parametrization'.

Simple conformable ε-parameterizations of realistic quark mixing angles is given by

$$\cos^2(2\theta^{\nu}_{12}) = f(-2\varepsilon), \; \cos^2(2\theta^{\nu}_{23}) = f(\varepsilon^2), \; \sin^2(2\theta^{\nu}_{13}) = f(\varepsilon). \quad (A6)$$

It coincides with solution (12) at $\varepsilon = 0$.

By geometric visualization, realistic solution (A6) means that the neutrino mixing vector $\mathbf{n}^{\nu}$, oriented at benchmark along the cube edge **c**, gets a little deviated from it by small not zero angle $2\theta^{\nu}_{13}$.

The ε-parameterization in (A6) breaks the residual symmetry of the bimaximal solution (12) and ensures agreement with data. Realistic neutrino mixing angles from solution (A6) are

$$\theta^{\nu}_{12} \cong 34°, \; \theta^{\nu}_{23} \cong 42.6°, \; \theta^{\nu}_{13} \cong 8.7°. \quad (A7)$$

From comparison with recent experimental data global analysis [**9**]

$$\theta^{\nu}_{12} = (33.7 \pm 1.1)°, \; \theta^{\nu}_{23} = (40.7 \pm 1.7)°, \; \theta^{\nu}_{13} = (8.8 \pm 0.4)° \quad (A8)$$

the solution (A7) for all three neutrino mixing angles agrees well with experimental data, also [7-9].

In terms of elemental electric charge (9), as alternative to ε parameter, realistic neutrino mixing angles are given by

$$\cos^2 2\theta_{sol} = f(-2e), \; \cos^2 2\theta_{atm} = f(e^2), \; \theta_{sol} \cong 33.9°, \; \theta_{atm} \cong 42.5°,$$

$$\sin^2 2\theta^{\nu}_{13} = f(-2e) + f(e^2), \; \theta^{\nu}_{13} \cong 11.5°. \quad (A9)$$

So, the elemental electric charge (9) as small parameter is almost as good for describing realistic neutrino mixing angles as the ε-parameter.

## 3. Parametrization of realistic quark mixing angles

Zero quark mixing pattern is the benchmark level for evaluation of realistic quark mixing angles. Symmetric equation for realistic not zero quark mixing angles is described by ε-parameterization of Eq. (7),

Realistic solution for quark mixing angles can be described by conformable parametrizations in analogy with neutrino ones

$$\sin^2 2\theta^q_{12} = f(2\varepsilon),\ \sin^2 2\theta^q_{23} = f(\varepsilon^2),\ \sin^2 2\theta^q_{13} = f(\varepsilon^4). \qquad (A10)$$

Note that the right side in the relations (A10) for the largest angle $\theta^q_{12}$ is different from the neutrino one in Eq. (A6) only by the sign of the ε-parameter.

Equations (A10) determine three realistic quark mixing angles

$$\theta^q_{12} \cong 13.05°,\ \theta^q_{23} \cong 2.36°,\ \theta^q_{13} \cong 0.19°. \qquad (A11)$$

By comparison with PDG data [2],

$$\theta^q_{12} = (13.02 \pm 0.04)°,\ \theta^q_{23} = (2.35 \pm 0.05)°,\ \theta^q_{13} = (0.20 \pm 0.01)°, \qquad (A12)$$

the agreement with predictions (A11) for all three quark mixing angles is remarkably good.

In terms of elemental electric charge (9), as alternative small parameter, realistic quark mixing angles are given by

$$\sin^2 2\theta^q_{12} = f(2e),\ \sin^2 2\theta^q_{23} = f(e^2),\ \sin^2 2\theta^q_{13} = f(e^4), \qquad (A13)$$

$$\theta_{12} \cong 13.38°,\ \theta_{23} \cong 2.46°,\ \theta_{13} \cong 0.21°. \qquad (A14)$$

They should be compared with (A11).and (A12).

## 4. Benchmark mixing pattern and the ε-parametrization

An important question should be answered of how are found the formulas for realistic quark (A10) and neutrino (A6) mixing angles. The answer is unexpectedly simple. All low energy dimensionless flavor quantities as small deviations from benchmark quark and neutrino ones are expressed through the epsilon-parameter, (8) or (9), via one universal function (A5) $f[y_i(\varepsilon)] = |y_i(\varepsilon) \exp y_i(\varepsilon)|$; $y_i(\varepsilon)$ are power functions that approximately fit empirical neutrino and quark mixing angles data. So, the problem is reduced to finding five power functions $y_i(\varepsilon)$ for three quark and two neutrino mixing angles. Luckily, data indications appear here surprisingly helpful. Instead of five different functions $y_i(\varepsilon)$, i = 1-5, we need only the easy to find by data fitting three simple ε-power ones: $y_1(\varepsilon) = 2\varepsilon$ for quark $\sin^2 2\theta^q_{12}$ in (A10) and $y_1(\varepsilon) = (-2\varepsilon)$ for neutrino $\cos^2 2\theta^\nu_{12}$ in (A6); $y_2(\varepsilon) = \varepsilon^2$ for both $\sin^2 2\theta^q_{23}$ in (A10) and $\cos^2 2\theta^\nu_{23}$ in (A6); and $\sin^2 2\theta^q_{13} = \varepsilon^4$ in (A10).

Important points here are empirically indicated good fitting qualities of universal ε-parameter (8) (or e -parameter (9) ) and f-function (A5) for the deviations of realistic neutrino and quark mixing angles from the primary benchmark mixing ones. The necessary condition for that fitting to happen is that benchmark neutrino mixing pattern must be bimaximal one (12) and the benchmark quark mixing pattern is zero one (18). Thus, *united* bimaximal neutrino and zero quark mixing patterns appear as nature's choice for elementary particle benchmark mixing.